\documentclass[useAMS,usenatbib]{mn2e}

\usepackage{txfonts,graphicx,amssymb,rotating}

\title[{[Ne\,{\sc ii}]} line profiles from disc winds]{[Ne\,{\sc ii}] emission line profiles from photoevaporative disc winds}

\author[Alexander]
  {R.D.~Alexander\thanks{email: rda@strw.leidenuniv.nl}\\Sterrewacht Leiden, Universiteit Leiden, Niels Bohrweg 2, 2300 RA, Leiden, the Netherlands}

\date{Accepted 2008 September 1. Received 2008 September 1; in original form 2008 August 12}

\begin{document}

\pagerange{\pageref{firstpage}--\pageref{lastpage}} \pubyear{2008}

\maketitle

\label{firstpage}

\begin{abstract}
I model profiles of the [Ne\,{\sc ii}] forbidden emission line at 12.81$\mu$m, emitted by photoevaporative winds from discs around young, solar-mass stars.  The predicted line luminosities ($\sim10^{-6}$L$_{\odot}$) are consistent with recent data, and the line profiles vary significantly with disc inclination.  Edge-on discs show broad (30--40\,km\,s$^{-1}$) double-peaked profiles, due to the rotation of the disc, while in face-on discs the structure of the wind results in a narrower line ($\simeq10$\,km\,s$^{-1}$) and a significant blue-shift (5--10\,km\,s$^{-1}$).  These results suggest that observations of  [Ne\,{\sc ii}] line profiles can provide a direct test of models of protoplanetary disc photoevaporation.
\end{abstract}



\section{Introduction}\label{sec:intro}
The manner in which gas is removed from discs around young stars has critical consequences for theories of planet formation, as the formation of gas-giant planets must precede the dissipation of gas discs.  Currently, models of protoplanetary disc evolution suggest that the most of the gas is removed by a combination of viscous accretion through the disc and photoevaporation by irradiation from the central star [e.g.,~\citet*{cc01,acp06b}; see also the recent reviews by \citet{dull_ppiv} and \citet{rda08}].  Such models agree reasonably well with current data, but the ionizing luminosities of T Tauri stars (required to drive the photoevaporation) are very poorly constrained.  \citet*{acp05} used archival data to estimate the chromospheric ionizing fluxes from a small sample of bright T Tauri stars, and found values in the range $\Phi \sim 10^{42}$--$10^{43}$ ionizing photons per second.  However, recent data suggests that this may over-estimate more typical values of $\Phi$ by 1--2 orders of magnitude \citep{herczeg07b}.  Photoevaporative flows have been observed clearly in ultracompact H\,{\sc ii} regions around massive stars \citep{holl94,lugo04}, and in the externally illuminated ``proplyds'' in the Orion nebula \citep*[e.g.][]{jhb98}.  However, the evidence for or against the existence of central star-driven photoevaporative winds from discs around solar-mass stars is rather tenuous.  \citet{font04} modelled the profiles of optical forbidden emission lines and found reasonable agreement with both the line fluxes and profiles obtained in the spectroscopic survey of \citet*{heg95}.  However, some discrepancies were also present (notably between the predicted and observed fluxes from [O\,{\sc i}]), and no further studies of such emission lines exist.

For several reasons, the [Ne\,{\sc ii}] fine-structure line at 12.81$\mu$m offers perhaps the best opportunity to confirm or deny the existence of a slow ($\sim10$\,km\,s$^{-1}$) ionized wind from a T Tauri disc.  The high ionization potential of neon (21.56eV) means that Ne$^+$ only exists close to T Tauri stars in photoionized gas, so forbidden emission lines from neon ions are unlikely to arise elsewhere in the T Tauri system.  The 12.81$\mu$m line also falls in an atmospheric window (unlike the [Ne\,{\sc iii}] line at 15.55$\mu$m), and can therefore be observed from the ground at echelle resolution.  Moreover, the continuum emission from the star-disc system is orders of magnitude weaker in the mid-infrared than in the optical, so the line-to-continuum ratio for [Ne\,{\sc ii}] can often be higher than for optical forbidden lines such as [N\,{\sc ii}] or [S\,{\sc ii}].  In addition, as will be shown, the critical density of the [Ne\,{\sc ii}] line is such that most of the emission arises in the ``launching region'' of the photoevaporative wind, making it an excellent tracer of disc photoevaporation.

Recently the {\it Spitzer Space Telescope} has detected the [Ne\,{\sc ii}] 12.81$\mu$m line towards more than 20 young, solar-mass stars \citep{pasc07,lahuis07,esp07}, and in most cases it is thought to arise in the star-disc system.  \citet{herczeg07a} observed the [Ne\,{\sc ii}] line from the nearby source TW Hya at high resolution ($\lambda/\Delta\lambda \approx 30,000$), and found that the line (with width $21\pm4$\,km\,s$^{-1}$) was narrower than lines that originate in the accretion funnel, but significantly broader than lines emitted from the disc at larger radii: this strongly suggests that the line originates in a hot disc atmosphere.  \citet*{gni07} and \citet{gh08} showed that X-ray heating of the disc can reproduce the observed line fluxes, but a photoevaporative disc wind should give rise to a similar line flux \citep{pasc07,gh08}.  While both ionization mechanisms (X-ray and UV) can likely explain the observed line fluxes, the velocity structure of a photoevaporative wind should result in a line profile that is distinct from that produced by X-ray irradiation of a bound disc atmosphere.  In this {\it Letter} I model the profiles of the [Ne\,{\sc ii}] 12.81$\mu$m line emitted by photoevaporative winds, and find that high-resolution spectroscopy can provide an unambiguous means of detecting such winds in T Tauri systems.


\section{Models}
\subsection{Hydrodynamic models}
In order to construct velocity profiles it is first necessary to compute the hydrodynamic structure of the wind.  Here two different cases are considered: the standard photoevaporative wind \citep{holl94,font04} and the ``direct'' wind, applicable to a disc with an inner hole that is optically thin to ionizing photons \citep*{acp06a}.  I use the {\sc zeus2d} hydrodynamics code \citep{sn92} to model the winds, following the approaches of \citet{font04} and \citet{acp06a} respectively.  I assume azimuthal and midplane symmetry, and therefore model the disc using a polar [$(r,\theta)$] grid covering the range $\theta=[0,\pi/2]$\footnote{Lower-case $r$ denotes spherical radius; upper-case $R$ cylindrical radius.}.  The grid cells are logarithmically-spaced in $r$ and linearly-spaced in $\theta$, with the numbers of radial ($N_r$) and polar ($N_{\theta}$) cells chosen so that the grid cells are approximately square throughout (i.e.~$\Delta r = r \Delta \theta$, see e.g., \citealt{bate02}).  The rotation option in  {\sc zeus2d}, which introduces a centrifugal pseudo-force, is turned on, and accelerations due to gravity are evaluated using only a point mass at the origin.  I adopt the van Leer (second-order) interpolation scheme, and the standard von Neumann \& Richtmyer form for the artificial viscosity (with $q_{\mathrm {visc}}=2.0$).  

\subsubsection{Standard case}
In the standard case the wind models follow the ``photoevaporative disc wind'' models of \citet{font04}.  These models make use of the results of \citet[][specifically the ``weak-wind'' case]{holl94}, who used detailed radiative transfer calculations to determine the structure of the ionized disc atmosphere.  The atmosphere is isothermal at $10^4$K, with a sound speed $c_{\mathrm s}=10$\,km\,s$^{-1}$.  The critical length-scale $R_{\mathrm g}$, known as the gravitational radius, is found by equating the sound speed of the ionized gas with the local (Keplerian) orbital speed, so 
\begin{equation}
R_{\mathrm g} = \frac{GM_*}{c_{\mathrm s}^2} \simeq 8.9 \, \left(\frac{M_*}{1\mathrm M_{\odot}}\right)\, \mathrm {AU} \, ,
\end{equation}
where $M_*$ is the mass of the central star.  The (number) density at the base of the ionized atmosphere, $n_0(R)$, determines the structure of the wind, and the value at $R_{\mathrm g}$ is
\begin{equation}\label{eq:base_den}
n_{\mathrm g} = C \left( \frac{3\Phi}{4\pi\alpha_{\mathrm B}R_{\mathrm g}^3}\right)^{1/2} \simeq 2.8\times10^4 \left(\frac{\Phi}{10^{41}\mathrm s^{-1}}\right)^{1/2} \left(\frac{M_*}{1\mathrm M_{\odot}}\right)^{-3/2}    \, .
\end{equation}
$\alpha_B=2.6\times10^{-13}$cm$^3$s$^{-1}$ is the Case B recombination coefficient for atomic hydrogen at $10^4$K \citep{allen}, and the constant $C$ was determined by the numerical calculations of \citet{holl94}.  Inside $R_{\mathrm g}$ the base density scales as $R^{-3/2}$, while outside $R_{\mathrm g}$ it scales as $R^{-5/2}$.  \citet{font04} adopt the following fitting form for the base density, which varies smoothly between the two power-laws:
\begin{equation}
n_0(R) = n_{\mathrm g} \left(\frac{2}{(R/R_{\mathrm g})^{15/2} + (R/R_{\mathrm g})^{25/2}}\right)^{1/5} \, .
\end{equation}
The disc atmosphere is given an isothermal equation of state, and the outer and inner radial boundaries are set as outflow boundaries.  The inner polar boundary (the $z$-axis) is a reflecting boundary, but the flow is not at all sensitive to the inner boundary conditions.  The base of the flow (i.e.~the disc midplane) is held constant at every time-step, using the density profile $n_0(R)$ from Equation \ref{eq:base_den}.  [The base of the flow is actually found several disc scale-heights above the midplane \citep{holl94}, but additional calculations show that this approximation has a negligible effect on the line profiles.]  The flow is thus assumed to be recombination-dominated (with negligible advection across the ionization front): previous studies suggest that this approximation is valid \citep[e.g.][]{acp06a}.  The base cells are given a Keplerian velocity in the orbital direction, and zero velocity in the radial direction.  The polar (vertical) velocity out of the base cells is not prescribed, but instead computed hydrodynamically: typically the launch velocity is $\simeq0.3$--0.4$c_{\mathrm s}$ \citep{font04}.  The models are computed in dimensionless units: the unit of length is $R_{\mathrm g}$, the unit of time is the orbital period at $R_{\mathrm g}$, and the density is normalized so that $n_{\mathrm g}=1$.  Initially the grid is filled with a uniform density of $10^{-8}n_{\mathrm g}$, and the model is evolved forward in time until a steady-state is reached.  In order to compute line profiles, the resulting density and velocity structures are scaled to physical units by choosing the parameters $\Phi$ and $M_*$.

\begin{figure}
\centering
       \resizebox{\hsize}{!}{
       \includegraphics[angle=90]{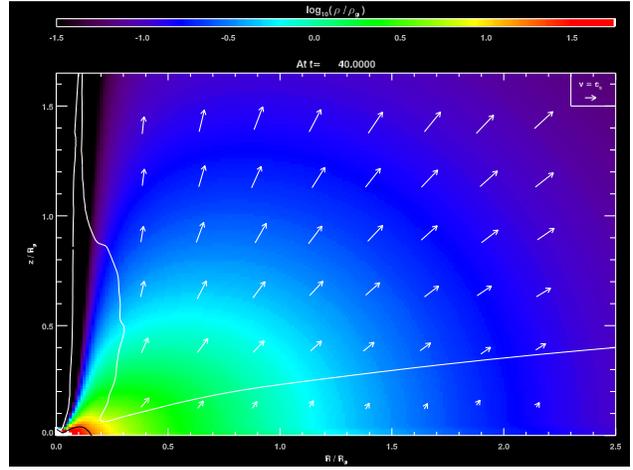}
       }
       \caption{Density and velocity structure of the steady-state photoevaporative flow.  $z=0$ corresponds to the base of the disc atmosphere; the underlying disc is not shown.  The plot shows the ``launching region'' of the wind; the grid extends to $r=20$R$_{\mathrm g}$.  The density is shown on the colour-scale, with velocity vectors plotted at regular intervals.  The white line shows the sonic surface, where the velocity in the $(r,\theta)$ plane is equal to the sound speed.  The black contour indicates where the local density is equal to the critical density for the [Ne\,{\sc ii}] line, for parameters $\Phi=10^{41}$s$^{-1}$ and $M_*=1$M$_{\odot}$.}     
           \label{fig:PDW_struc}
\end{figure}

I use a grid with $r_{\mathrm {in}} = 0.03$R$_{\mathrm g}$, $r_{\mathrm {out}} = 20$R$_{\mathrm g}$, $N_r = 246$ and $N_{\theta}=59$.  As one would expect there is excellent agreement with the results of \citet{font04}, with the location of the sonic surface and the integrated mass-loss rate agreeing with their result to better than 1\% accuracy.  Choosing different locations for the boundaries ($r_{\mathrm {in}} = 0.01$R$_{\mathrm g}$; $r_{\mathrm {out}} = 50$R$_{\mathrm g}$) has a negligible effect on the structure of the flow and affects the computed line fluxes and profiles only at the percent level.  A numerical convergence test (using twice the grid resolution in both dimensions) suggests that the flow solution is correct to 1--2\% accuracy.  A steady-state is reached after $t=5$--10 time units; the subsequent line profile modelling makes use of the density and velocity fields at $t=40$.  The steady-state flow solution is shown in Fig.\ref{fig:PDW_struc}.

\subsubsection{Discs with holes}
For the case of a disc with an optically thin inner cavity, the model of \citet{acp06a} is used.  In this case the calculation is not scale-free (because of the ionization/recombination balance calculation), so physical units are used.  In addition, because the pressure scale-height of the cold disc must be resolved into several grid cells, somewhat higher resolution is required than in the standard case (where the underlying cold disc is treated as a boundary condition).  I adopt model parameters of $M_* = 1$M$_{\odot}$ and $\Phi=10^{41}$s$^{-1}$, and grid parameters $r_{\mathrm {in}} = 5$AU, $r_{\mathrm {out}} = 200$AU, $N_r = 360$ and $N_{\theta}=153$.  The outer radius is chosen to be approximately equal to that used in the standard case when scaled to physical units (for $M_* = 1$M$_{\odot}$), but the results are not sensitive to the exact location of the boundary.  The initial disc follows an power-law surface density profile $\Sigma \propto R^{-1}$, truncated exponentially at some inner radius, and is locally isothermal in the vertical direction with aspect $H/R=0.05$.  The total disc mass is $0.001$M$_{\odot}$ (consistent with the low disc mass expected during photoevaporative clearing), but the resulting line profiles are not sensitive to this mass.  The model is allowed to evolve to a ``quasi-steady'' state\footnote{The flow in this case is only ``quasi-steady'' because the mass reservoir is finite.  This state is reached in $\lesssim 10^3$yr, which is much less than the time-scale for disc clearing ($\sim10^5$yr).  As a result, aside from initial transients the flow solution does not change significantly over the duration of the simulations.} before computing line profiles, and considers a inner hole of radius 9AU ($\simeq$R$_{\mathrm g}$).  The quasi-steady state is reached after 1--2 outer orbital periods; the line profiles are calculated from the density and velocity fields at $t=4000$yr.

\subsection{Line profiles}\label{sec:profiles}
In order to compute line profiles from the hydrodynamic models, it is first necessary to construct three-dimensional density and velocity fields from the two-dimensional simulations.  I assume reflective symmetry at the disc midplane in order to extend the coordinate range to $\theta=[0,\pi]$, and azimuthal symmetry around the polar axis.  The resulting grid has $N_r \times 2N_{\theta} \times N_{\phi}$ cells: an azimuthal resolution of $N_{\phi}=120$ is adopted, but the exact value of $N_{\phi}$ has no significant effect on the results.

The line luminosity $L$ at a given velocity $v$ is computed as
\begin{eqnarray}
L(v) = \frac{1}{\sqrt{2\pi}v_{\mathrm {th}}} \int \exp\left(-\frac{[v-v_{\mathrm {los}}(\mathbf r)]^2}{2v_{\mathrm {th}}^2}\right) \, Ab_{\mathrm {Ne}} \, X_{\mathrm {II}}
\nonumber \\
\times \, n_e(\mathbf r) \, P_u \, A_{ul} \, h\nu_{ul} \,  dV \, ,
\end{eqnarray}
with the integral evaluated by direct summation over the entire extent of the grid.  Here $v_{\mathrm {los}}(\mathbf r)$ is the line-of-sight component of the local gas velocity vector,  computed for a specified inclination angle $i$.  ($i=\pi/2$ corresponds to a disc viewed edge-on; $i=0$ face-on.)  $n_e(\mathbf r)$ is the local number density of the gas (hydrogen is assumed to be fully ionized), and $h\nu_{ul}$ is the energy of the emitted photons.  The Doppler broadening term depends on the thermal velocity of the emitting atoms
\begin{equation}
v_{\mathrm {th}} = c_{\mathrm s}\sqrt{m_{\mathrm H}/m_{\mathrm {Ne}}} \, ,
\end{equation}
where $m_{\mathrm H}$ and $m_{\mathrm {Ne}}$ are the masses of hydrogen and neon atoms respectively.  I adopt the standard solar value for the abundance of neon,  $Ab_{\mathrm {Ne}}=1.0\times10^{-4}$, and the Einstein coefficient of the transition is $A_{ul} = 8.39\times 10^{-3}$s$^{-1}$ \citep{mendoza83,gni07}.  $X_{\mathrm {II}}$ is the fraction of neon that exists as Ne$^+$, discussed below.  Following \citet{gni07}, the excitation fraction of the upper state $P_u$ is computed as
\begin{equation}\label{eq:P_u}
P_u = \frac{1}{2 C_{ul} \exp(-T_{ul}/T) +1} \, ,
\end{equation}
where the excitation temperature $T_{ul} = 1122.8$K, and the gas temperature $T=10,000$K.  $C_{ul}$ expresses the departure from a thermal level population, and is defined as
\begin{equation}
C_{ul} = 1 + \frac{n_{\mathrm {cr}}}{n_e(\mathbf r)} \, ,
\end{equation}
where the critical density $n_{\mathrm {cr}} = 5\times10^5$cm$^{-3}$.  I assume that the wind is optically thin to the emitted line, but that the disc absorbs 100\% of the line flux if the midplane intercepts the line-of-sight to the observer.  This approach neglects line emission from partially ionized gas in the ionization front, but analytic estimates suggest this contributes to the total line luminosity only at the percent level.

The factor $X_{\mathrm {II}}$ is determined by ionization/recombination balance.  The first ionization potential of atomic neon is 21.56eV, so in the $10^4$K disc atmosphere collisional ionization of neon is negligible.  Instead, neon is photo-ionized, either by UV photons with energies greater than 21.56eV or by high-energy X-rays (via the Auger mechanism).  The low line-of-sight column density in the wind region ($10^{17}$--$10^{18}$cm$^{-2}$) suggests that UV ionization is more efficient than ionization by X-rays, and I neglect X-ray ionization here.  The second ionization potential of neon is 41.0eV, so the relative abundances of Ne, Ne$^+$ and Ne$^{2+}$ are determined by the spectral slope of the incident UV radiation field.  Little is known about the details of the incident radiation field, but comparison to Galactic H\,{\sc ii} regions suggests that $X_{\mathrm {II}}$ likely lies in the range 0.1--1.0 \citep[e.g.][]{rubin91}.  In the models almost all the ionized gas is below the critical density, so the line flux per unit volume is simply proportional to $X_{\mathrm {II}}$.  Moreover, the bulk of the line flux arises from a fairly limited range in radius ($\simeq0.1$--2.0$R_{\mathrm g}$; see discussion in Section \ref{sec:results} below), and $X_{\mathrm {II}}$ is unlikely to vary significantly over this range.  Ionization balance therefore has a negligible effect on the shape of the line profile, but the integrated line flux scales linearly with $X_{\mathrm {II}}$.  In the absence of any knowledge of the incident spectrum of radiation I adopt a constant value of $X_{\mathrm {II}}=1/3$, noting that this introduces an uncertainly of a factor of $\sim3$ in the derived line luminosities.


\section{Results}\label{sec:results}
\begin{table}
 \centering
  \begin{tabular}{cccccc}
  \hline
Inclination & Line flux &  Peak  &  FWHM & Peak & FWHM \\
& $10^{-6}$L$_{\odot}$ & \multicolumn{2}{c}{Model} & \multicolumn{2}{c}{$\lambda/\Delta\lambda=30$,000} \\ \hline

\multicolumn{6}{c}{Standard model} \\
$\pi/2$ & 3.0 & $\pm$10.6 & 33.5 & $\pm$7.6 & 38.1 \\
$\pi/4$ & 1.5 & $-$12.1/$+$0.9 & 26.3 & $-$7.4 & 28.7  \\
0 & 1.5 & $-$6.4 & 9.5 & $-$6.6 & 16.8 \\
\\
\multicolumn{6}{c}{Disc with $R_{\mathrm g}$-sized inner hole} \\
$\pi/2$ & 5.4 & $\pm$3.1 & 17.1 & 0.0 & 19.3 \\
$\pi/4$ & 2.7 & $-$4.8 & 13.3 & $-$6.3 & 18.1 \\
0 & 2.7 & $-$7.9 & 10.5 & $-$8.9 & 17.0 \\
\hline
\end{tabular}
  \caption{Details of the line profiles shown in Figs.\ref{fig:line_disc} \& \ref{fig:line_10au}, computed for a stellar mass $M_*=1$M$_{\odot}$ and an ionizing flux $\Phi=10^{41}$s$^{-1}$.  The columns labelled ``Peak'' show the velocities at which the line fluxes are maximum, while the full widths of the lines at half the maximum flux are labelled ``FWHM''.  ``Model'' denotes the values for the theoretical line profiles, while ``$\lambda/\Delta\lambda=30$,000'' indicates the values when observed at the typical resolution of an echelle spectrograph.  Velocities with positive and negative values indicate that the profile is double-peaked.  All velocities are in \,km\,s$^{-1}$; the ambient disc gas has zero velocity.}
  \label{tab:results}
\end{table}

\begin{figure}
\centering
       \resizebox{\hsize}{!}{
       \includegraphics[angle=270]{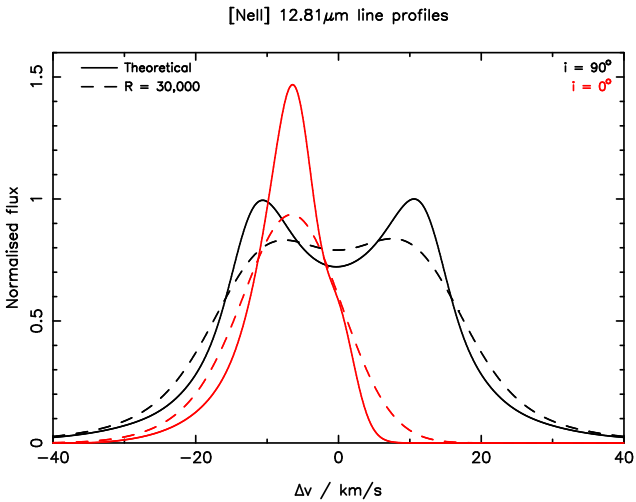}
       }
       \caption{Line profiles for the photoevaporative disc wind.  The black lines show the profiles when the disc is viewed edge-on ($i=\pi/2$), the red lines the profiles for a face-on disc ($i=0$).  The dashed lines show the profiles degraded to typical echelle resolution ($\lambda/\Delta\lambda=30,000$).  The flux per unit velocity interval is normalised to the peak value in the edge-on case.  When viewed edge-on the line is broad and double-peaked, reflecting the large, 10--20\,km\,s$^{-1}$, orbital velocity in the emitting region.  When viewed face-on the disc rotation is perpendicular to the line-of-sight, and instead the vertical component of the wind results in the line being blue-shifted by 5--10\,km\,s$^{-1}$.}     
           \label{fig:line_disc}
\end{figure}

The results of the line-profile modelling are shown in Table \ref{tab:results}.  Two disc models are presented (standard, 9AU hole) for model parameters $M_*=1$M$_{\odot}$ and an $\Phi=10^{41}$s$^{-1}$, and for each model the line profile was computed for inclinations of $i=0$ (face-on), $i=\pi/4$ and $i=\pi/2$ (edge-on).  In addition, ``realistic'' line profiles, at the typical resolution of an echelle spectrograph, were computed by convolving the line profiles with a Gaussian profile with half-width $\sigma=5$\,km\,s$^{-1}$: the effective resolution of these profiles is $\lambda/\Delta\lambda=$30,000.

The line profiles for the standard case are shown in Fig.\ref{fig:line_disc}: for clarity, only the edge-on and face-on cases are shown.  When viewed edge-on the line is broad (30--40\,km\,s$^{-1}$) and double-peaked, while face-on inclination results in a line that is narrower (10\,km\,s$^{-1}$) and slightly blue-shifted (6--7\,km\,s$^{-1}$).  At resolution $\lambda/\Delta\lambda=30,000$ the blue-shift is comparable for all inclinations $i \lesssim \pi/4$ (see Table \ref{tab:results}), as any double-peaked structure in the line is not well-resolved.  These line profiles can be easily understood by considering the flow structure.  At large radii the wind is essentially isothermal and spherically symmetric, and is analogous to a Parker wind solution.  The streamlines are close to radial, and the density along streamlines drops as $n_e \propto r^{-2}$.  As the density is sub-critical the line flux per unit volume scales $\propto n_e^2 \propto r^{-4}$, so the total line luminosity (integrated over volume) is dominated by the emission from small radii.  Inspection of the mass-loss profile (see, e.g., Fig.7 of \citealt{font04}) shows that $>80$\% of the mass-loss comes from $<2R_{\mathrm g}$, and the density structure is such that the critical density is reached only at radii $\lesssim 0.1R_{\mathrm g}$ (see Fig.\ref{fig:PDW_struc}).  This highlights why the [Ne\,{\sc ii}] 12.81$\mu$m line is an ideal probe of disc photoevaporation: lines with higher critical densities will likely be dominated by emission from high density regions very close to the star, while lines with lower critical densities are insensitive to the launching region of the wind.  In the launching region the rotation speed of the disc is greater than the wind or thermal speeds (as $v_{\mathrm K} = c_{\mathrm s}$ at $R=R_{\mathrm g}$), so when viewed edge-on the line profile is dominated by Keplerian rotation and shows a pronounced double-peaked structure.  When viewed face-on the rotation is entirely in the plane of the sky, and instead the vertical component of the wind velocity results in a blue-shifted profile.  In this case the profile is slightly asymmetric, with a blue ``tail'' due to high-velocity, low-density gas at large $r$.  In all cases the line-widths are significantly larger than expected from thermal broadening alone.  This occurs because the streamlines are almost radial at large radius. Consequently the dispersion in the line-of-sight component of the flow velocity is always larger than the thermal line-width, resulting in broader than thermal line-widths even for face-on discs.
\begin{figure}
\centering
       \resizebox{\hsize}{!}{
       \includegraphics[angle=270]{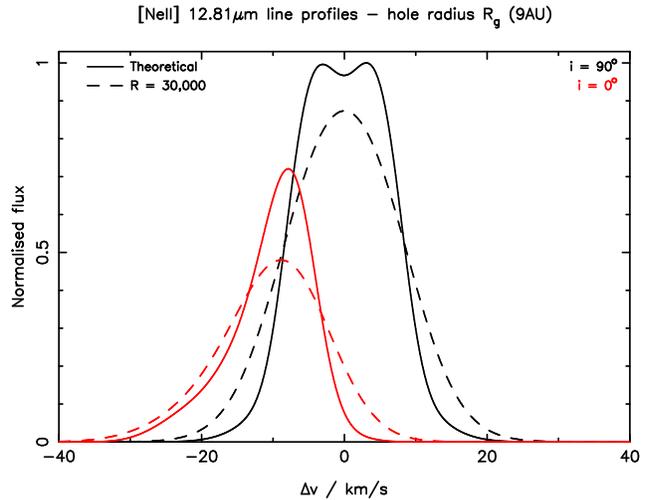}
       }
       \caption{As Fig.\ref{fig:line_disc}, but for a disc with a 9AU inner hole.  The face-on line profile is similar to that seen in the standard case, but in the edge-on case the absence of the inner disc results in a narrower line, with a less pronounced double-peaked profile.}     
           \label{fig:line_10au}
\end{figure}

In the case of a disc with a cleared central hole, the line profile when viewed edge-on is markedly different from the standard case (see Fig.\ref{fig:line_10au}).  The removal of gas with high rotation speeds results in a profile that is much less strongly double-peaked, and significantly narrower (15--20\,km\,s$^{-1}$).  Still larger hole sizes result in even less pronounced double-peaked profiles, and for hole sizes $\gtrsim 5$R$_{\mathrm g}$ there is no evidence of a double-peaked profile.  The profile in the face-on case is similar to that seen in the standard model: fairly narrow ($\simeq10$\,km\,s$^{-1}$), slightly asymmetric, and blue-shifted (7--9\,km\,s$^{-1}$), and this general shape is seen irrespective of the size of the central hole.  The integrated line flux is somewhat larger than in the standard case (by a factor of $\simeq 2$), due to the higher density in the wind launching region.

The predicted line luminosities are a few $\times10^{-6}$L$_{\odot}$.  The exact line fluxes are likely only accurate to within a factor of $\sim3$, as discussed in Section \ref{sec:profiles} above, but the predicted line strengths are comparable to those predicted by previous calculations, both from UV ionization \citep{gh08} and X-ray ionization \citep{gni07}.  The emitted line fluxes scale approximately linearly with $\Phi$, and the shape of the line profile is largely independent of the value of $\Phi$.  This can be understood from Equations \ref{eq:base_den} \& \ref{eq:P_u}: the line flux scales $\propto n_e^2$ as long as $n_e \ll n_{\mathrm {cr}}$, and the density of the ionized gas scales $\propto \Phi^{1/2}$.  We see from Fig.\ref{fig:PDW_struc} that almost all of the gas in the emitting region is below the critical density for [Ne\,{\sc ii}], and additional calculations suggest that this holds as long as $\Phi\lesssim10^{43}$s$^{-1}$.  Higher still ionizing fluxes will result in the critical density being exceeded, making the [Ne\,{\sc ii}] line less sensitive to the launching region of the wind, but this is unlikely to be relevant to discs around T Tauri stars.

\section{Discussion}
In general, these results compare favourably with the previous modelling of \citet{font04}.  These authors modelled the emission from optical forbidden lines ([S\,{\sc ii}] and [N\,{\sc ii}]), and for face-on inclinations found that the line profiles were blue-shifted by $\simeq10$\,km\,s$^{-1}$.  The blue-shift decreased with increasing inclination angle, and in the edge-on discs the lines were centred on zero velocity.  They did not, however, find any double-peaked lines, even in the edge-on case.  The reasons for this are not entirely clear, but it is likely attributable to the different critical densities of the lines considered.  The critical densities of the [S\,{\sc ii}] and [N\,{\sc ii}] lines are lower that that of the [Ne\,{\sc ii}] 12.81$\mu$m line, so these lines are dominated by emission from larger radii, where the rotation speed of the disc is smaller.  Consequently, as in the case of a disc with a large inner hole, the lines do not appear double-peaked.  

So far I have neglected the effects of X-ray ionization in producing [Ne\,{\sc ii}] emission, but when discussing observations this cannot be ignored.  \citet{gni07} modelled the effects of X-ray irradiation in detail \citep[see also][]{gh08}.  Unlike in the UV case, the X-ray luminosities of T Tauri stars are well-known, and the predicted [Ne\,{\sc ii}] emission from X-ray ionized gas is comparable in strength to the emission from the photoevaporative wind ($\sim 10^{-6}$L$_{\odot}$).  X-ray heating likely creates a hot, but static, disc atmosphere, and in real systems the emission from this atmosphere will be added to the line profiles modelled here.  The line profiles from such a bound atmosphere have not been modelled in detail, but their general shape was discussed by \citet{gni07}.  They find that emission is dominated by gas at small radii $(\lesssim15$AU), with most of the emission arising at radii where the rotational speed of the disc gas is $\sim10$\,km\,s$^{-1}$.  The gas temperature in this region is 1000--5000K, resulting in an intrinsic (Doppler) line-width of $\sim 1$\,km\,s$^{-1}$.  The line should therefore be double-peaked when viewed edge-on, with a profile similar to that predicted here.  In the face-on case, however, the profile should be different: narrow ($\sim 1$\,km\,s$^{-1}$), and centred on zero velocity.  The broader ($\sim 10$\,km\,s$^{-1}$) line-width predicted here is unlikely to be matched by a static disc atmosphere unless any turbulence in the disc is highly supersonic \citep[which is generally not the case in protoplanetary discs, e.g.,][]{bh98}, or unless the emission originates very close to the star ($\lesssim 1$AU).  In addition, a static atmosphere results in line emission centred on zero velocity, distinct from the blue-shifted profile of the photoevaporative wind.  Critical to observing this blue-shift are the relative line fluxes from the static atmosphere and the wind.  If these are comparable, as predicted for $L_X\sim10^{30}$erg s$^{-1}$ and $\Phi \sim 10^{41}$s$^{-1}$, then the diagnostic blue-shift will likely be 2--5\,km\,s$^{-1}$.

Recently \citet{herczeg07a} observed the [Ne\,{\sc ii}] 12.81$\mu$m line from the nearby face-on ($i\simeq7^{\circ}$) disc TW Hya, at a resolution of $\lambda/\Delta\lambda \approx 30,000$.  The emission was unresolved at an effective spatial resolution of $\sim40$AU, suggesting that it arises within this distance of the central star.  \citet{herczeg07a} measured a line-width (FWHM) of $21\pm4$\,km\,s$^{-1}$, centred on $-2\pm3$\,km\,s$^{-1}$.  The blue-shift is not statistically significant, in part because of the limitations of the wavelength calibration, but the line width is rather larger than expected from a static disc atmosphere alone.  In addition there was some evidence for asymmetry in the profile, with an enhanced flux on the blue side of the line, but not at a statistically significant level.  The observed profile is consistent with emission from a photoevaporative wind (centred on $\simeq -5$--10\,km\,s$^{-1}$) combined with the emission from a heated disc atmosphere (centred on zero velocity).  Further such observations are expected in the near future, and in this {\it Letter} I have shown that high-resolution observations of the [Ne\,{\sc ii}] 12.81$\mu$m line can provide a critical test of models of disc photoevaporation.  Detection of blue-shifted [Ne\,{\sc ii}] emission would provide unambiguous evidence of a photoevaporative wind, and observations of these line profiles may represent the most readily observable diagnostic of central star-driven disc photoevaporation.

\section*{Acknowledgements}
I am grateful for a number useful discussions with Greg Herczeg, Brent Groves and Ilaria Pascucci.  I also thank Cathie Clarke and Ewine van Dishoeck for comments on the manuscript, and the referee, Will Henney, for a thoughtful and insightful report.  This work was supported by the Netherlands Organisation for Scientific Research (NWO) through VIDI grants 639.042.404 and 639.042.607.   


\label{lastpage}

\end{document}